\documentclass[twocolumn,10pt]{IEEEtran}

\usepackage{amsmath,amssymb,amsthm}
\usepackage{graphicx}
\usepackage{xcolor}
\usepackage{booktabs}
\usepackage[acronym]{glossaries}

\newacronym{mimo}{MIMO}{multiple-input multiple-output}
\newacronym{rf}{RF}{radio frequency}
\newacronym{milac}{MiLAC}{microwave linear analog computer}
\newacronym{sim}{SIM}{stacked intelligent metasurface}
\newacronym{svd}{SVD}{singular value decomposition}
\newacronym{ris}{RIS}{reconfigurable intelligent surface}
\newacronym{adc}{ADC}{analog-to-digital converter}
\newacronym{dac}{DAC}{digital-to-analog converter}
\newacronym{r-zfbf}{R-ZFBF}{regularized zero-forcing beamforming}
\newacronym{mmse}{MMSE}{minimum mean square error}
\newacronym{lmmse}{LMMSE}{linear minimum mean square error}
\newacronym{qam}{QAM}{quadrature amplitude modulation}
\newacronym{nn}{NN}{neural network}
\newacronym{pcb}{PCB}{printed circuit board}
\newacronym{dft}{DFT}{discrete Fourier transform}
\newacronym{espar}{ESPAR}{electronically steerable parasitic array radiator}
\newacronym{lma}{LMA}{load modulated array}
\newacronym{dma}{DMA}{dynamic metasurface antenna}
\newacronym{bd-ris}{BD-RIS}{beyond diagonal RIS}
\newacronym{dsa}{DSA}{dynamic scattering array}

\begin{document}
\bstctlcite{BSTcontrol}

\title{Microwave Linear Analog Computers (MiLACs)\\for Communications: Opportunities and Challenges}

\author{Matteo~Nerini,~\IEEEmembership{Senior~Member,~IEEE}, and
        Bruno~Clerckx,~\IEEEmembership{Fellow,~IEEE}

\thanks{This work has been supported in part by UKRI under Grant EP/Y004086/1, EP/X040569/1, EP/Y037197/1, EP/X04047X/1, EP/Y037243/1.}
\thanks{The authors are with the Department of Electrical and Electronic Engineering, Imperial College London, SW7 2AZ London, U.K. (e-mail: m.nerini20@imperial.ac.uk; b.clerckx@imperial.ac.uk).}}

\maketitle

\begin{abstract}
Future wireless systems will require ever larger antenna arrays and heavier signal processing, making conventional digital \gls{mimo} architectures difficult to scale.
In this paper, we show that a possible solution is to offload part of the processing from the digital to the analog domain.
This can be done through linear microwave networks designed to compute directly using the communication signals at \gls{rf}.
These networks, denoted as \glspl{milac}, can perform useful matrix operations instantly through wave propagation.
Remarkably, although \glspl{milac} are linear, the output signals can depend nonlinearly on the tunable parameters of the network, enabling the computation of operations beyond simple linear transforms.
In particular, \glspl{milac} can realize matrix inversion and pseudo-inversion with complexity scaling quadratically with matrix size, rather than cubically, which is essential in zero-forcing beamforming.
We then review how \gls{milac}-aided \gls{mimo} architectures can reduce the number of \gls{rf} chains, relax the resolution requirements on \glspl{dac} and \glspl{adc}, and decrease the beamforming complexity.
We finally discuss the main challenges related to \gls{milac} and promising directions for future research.
\end{abstract}

\glsresetall


\section{Introduction}

Next-generation wireless networks are expected to deliver far higher data rates, broader coverage, lower latency, and connectivity for massive numbers of devices.
In 6G, these demands are being addressed in the 7-24~GHz range, where propagation remains favorable, with arrays of hundreds or even thousands of antennas to provide the required spectral efficiency.
This creates the challenge of designing \gls{mimo} base stations that can scale to such dimensions.
A critical bottleneck in modern base stations is their heavy reliance on digital processing.
Digital \gls{mimo} precoding and combining require a large number of \gls{rf} chains, each including expensive and power-hungry components such as \glspl{dac} and \glspl{adc}.
In addition, the involved matrix operations become computationally prohibitive at very large scales.
A promising way forward is therefore to offload part of the processing from the digital baseband domain to the analog \gls{rf} domain through analog computing strategies \cite{del26}.

Analog computing refers to performing mathematical operations using physical quantities whose behavior is \textit{analogous} to the mathematical quantities they represent.
A promising form of analog computing is wave-domain (or wave-based) computing \cite{del26}.
Here, the physical quantities exploited to compute are waves, whether mechanical (e.g., sound waves) or electromagnetic (e.g., microwaves or light).
In wireless communications, the natural choice of waves for analog computing is microwaves, i.e., in the 0.3-300~GHz range, since they are already used for information transmission.
Moreover, microwaves commonly propagate through linear media, such as air, or networks of linear components, such as transmission lines, phase shifters, capacitors, and inductors.
This makes linear microwave networks the natural class of analog computers for future wireless communications.

In this paper, we refer to any linear microwave network used to perform mathematical operations in the analog domain as a \gls{milac}, and review its potential for wireless communications \cite{ner25-1,ner25-2}.
First, we present a general model for analog computing with linear microwave networks and illustrate several representative \gls{milac} implementations.
A \gls{milac} may be fixed or reconfigurable, may compute over-the-air or on a \gls{pcb}, and can realize matrix operations directly through wave propagation.
Interestingly, the operations that can be computed with \glspl{milac} are not limited to linear transformations when the input data is encoded into the tunable parameters of the microwave network.
For instance, matrix inversion and pseudo-inversion can also be computed in the analog domain.
Second, we show how \glspl{milac} can realize beamforming in the analog domain by reviewing \gls{milac}-aided transmitting and receiving architectures, including advanced architectures that perform as fully digital beamforming.
Remarkably, \gls{milac} offers major benefits in minimizing the required number of \gls{rf} chains, the resolution of \glspl{dac} and \glspl{adc}, and the computational complexity of beamforming.
Third, we highlight open research problems, including modeling the hardware impairments, channel estimation, wideband operation, and developing \gls{milac} architectures with low hardware complexity.

\begin{figure}[t]
\centering
\includegraphics[width=0.32\textwidth]{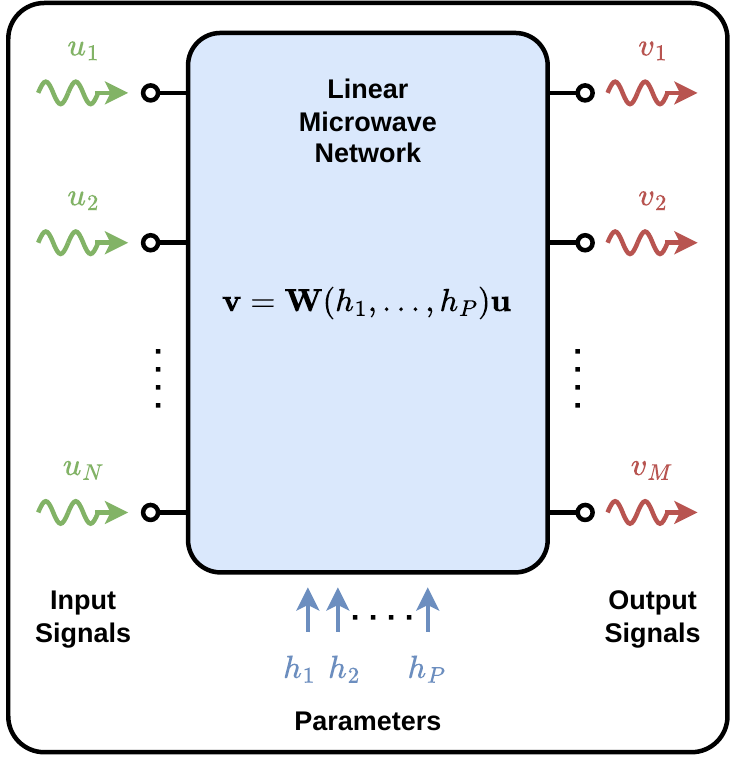}
\caption{A MiLAC whose linear microwave network has $N$ input ports, $M$ output ports, and $P$ tunable parameters.}
\label{fig:milac}
\end{figure}

\section{Analog Computing with\\Linear Microwave Networks}

In this section, we show how linear microwave networks can act as analog computers, referred to as \glspl{milac}.
We first present a general narrowband model for \glspl{milac}.
We then illustrate representative examples of microwave networks that can be used to compute.
Finally, we discuss the computational capabilities of these networks and the complexity reductions they offer compared to digital computing.

\subsection{A General Narrowband Model of MiLAC}

A \gls{milac} is a microwave network composed of linear components that is used to perform a computation.
To highlight its computational capabilities, we focus on the narrowband case.
A narrowband \gls{milac} processes sinusoidal microwave signals and, because of its linearity, produces sinusoidal signals as outputs at the same frequency.
As illustrated in Fig.~\ref{fig:milac}, it can be modeled as a multi-port microwave network receiving the input signals on $N$ ports and returning the outputs on $M$ ports.
Because the network is linear, each output is a linear combination of the inputs.
Let $\mathbf{u}\in\mathbb{C}^{N\times 1}$ and $\mathbf{v}\in\mathbb{C}^{M\times 1}$ denote the complex baseband representations of the input and output signals, respectively.
The input-output relationship can then be written as $\mathbf{v}=\mathbf{W}\mathbf{u}$, where $\mathbf{W}\in\mathbb{C}^{M\times N}$ is a matrix depending on the microwave network.

The microwave network can be either fixed or reconfigurable.
If the microwave network is fixed, the \gls{milac} can only compute a single linear transformation $\mathbf{W}$, while the input vector $\mathbf{u}$ can vary from one use to another according to the input signals.
This operation corresponds to a matrix-vector product, which would require $\mathcal{O}(MN)$ operations in the digital domain.
By contrast, in a \gls{milac}, the computation is carried out directly by wave propagation through the network and is completed within the physical propagation delay.
Since microwave signals travel in the network at a speed close to that of light, this delay is extremely small, so the computation can be regarded as instantaneous for engineering purposes.

The microwave network can also be reconfigurable, i.e., it can include tunable components such as tunable phase shifters, PIN diodes, or varactors.
In this case, the behavior of the \gls{milac} depends on the configuration of these components.
More specifically, when the network is controlled by $P$ tunable parameters $h_1,\ldots,h_P$, the resulting linear transformation depends on their configuration, and the output can be written as $\mathbf{v}=\mathbf{W}(h_1,\ldots,h_P)\mathbf{u}$.
This input-output relationship leads to two observations.
First, a reconfigurable \gls{milac} can compute different linear transformations of the input signals $\mathbf{W}(h_1,\ldots,h_P)$ depending on the values of the tunable parameters.
Second, although the output of a reconfigurable \gls{milac} remains linear in the input vector $\mathbf{u}$, it can depend nonlinearly on the tunable parameters $h_1,\ldots,h_P$.
This is a phenomenon known as structural nonlinearity \cite{del26}.
Hence, a linear microwave network can compute nonlinear functions with respect to its tunable parameters, which significantly broadens its computational capabilities.

\subsection{Examples of Linear Microwave Networks}
\label{sec:6}

\begin{figure*}[t]
\centering
\includegraphics[width=0.64\textwidth]{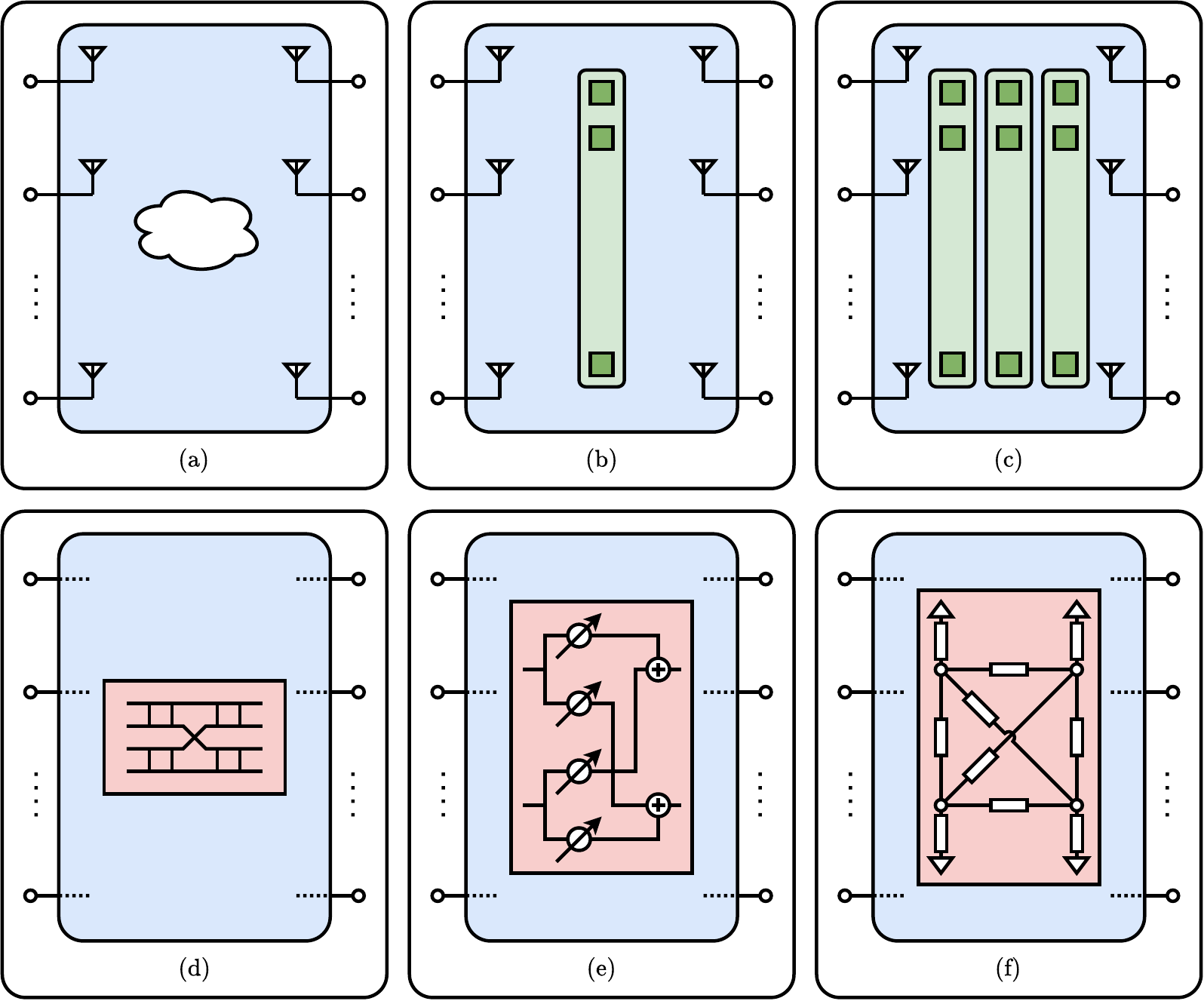}
\caption{Six examples of linear microwave networks that can implement a MiLAC: (a) a wireless channel, (b) a RIS-aided wireless channel, (c) a SIM-aided wireless channel, (d) a microstrip circuit, (e) a network of power dividers, phase shifters, and power combiners, and (f) a network of tunable impedance components.
In (a)-(c), the computation occurs over-the-air, while in (d)-(f), it occurs on-the-board.
In (a) and (d), the microwave network is not reconfigurable, while in the other cases it is reconfigurable and depends on tunable parameters.}
\label{fig:6}
\end{figure*}

\begin{figure}[t]
\centering
\includegraphics[width=0.48\textwidth]{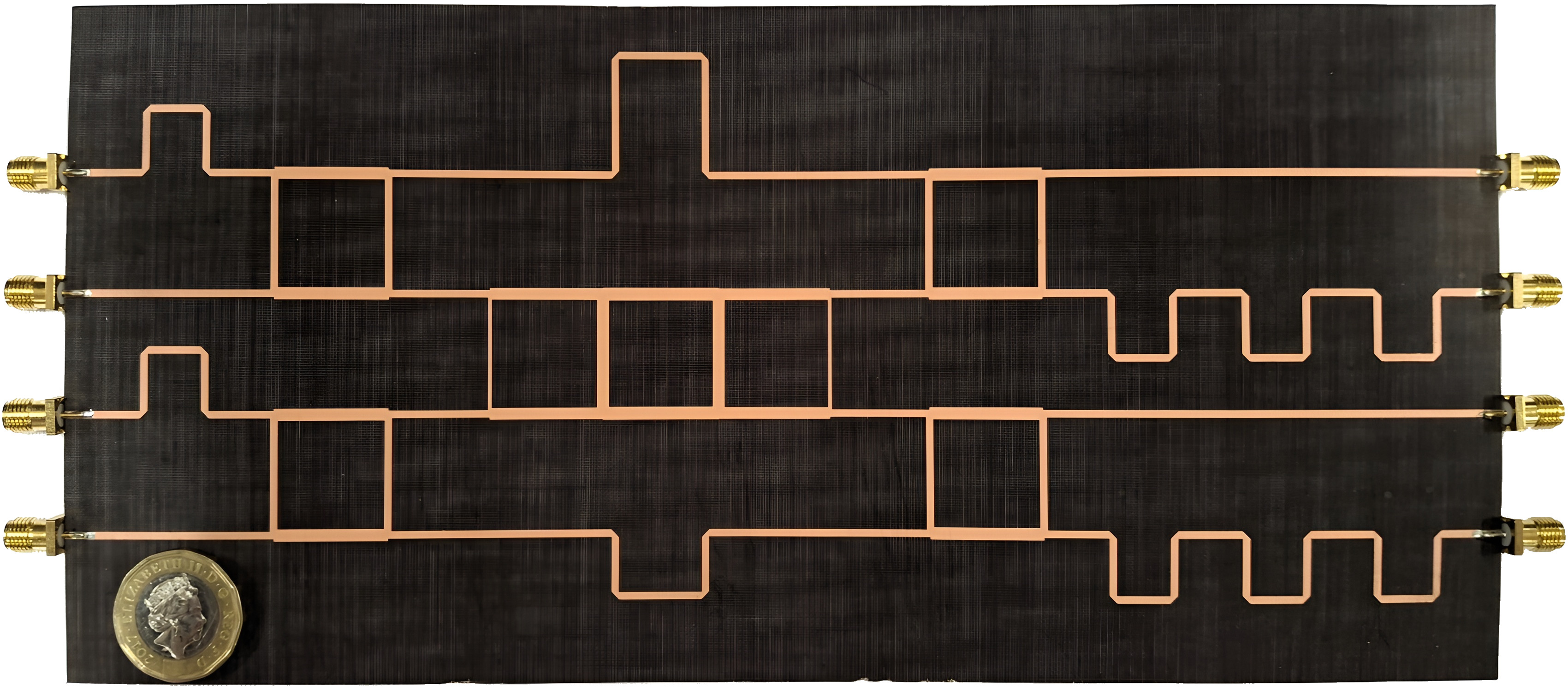}
\caption{A MiLAC fabricated in microstrip technology that computes the $4\times 4$ DFT operating at 2.4~GHz.}
\label{fig:dft}
\end{figure}

To make the discussion more concrete, we report six examples of linear microwave networks that can implement a \gls{milac} in Fig.~\ref{fig:6}.
The simplest example is a wireless channel, whose linearity can be exploited to perform computation.
As shown in Fig.~\ref{fig:6}(a), the input and output ports of the \gls{milac} are connected to transmitting and receiving antenna arrays, respectively, while the microwave network is the propagation channel between them.
Using a wireless channel in this way is commonly known as over-the-air computing.
In this case, the computed linear transformation is determined by the propagation environment and is therefore fixed for a given channel realization.

To enable the computation of different linear transformations, reconfigurability can be introduced into the wireless channel.
A way to achieve this is to deploy surfaces with reconfigurable reflective or transmissive properties, known as \glspl{ris} \cite{wu21}, within the propagation environment, as illustrated in Fig.~\ref{fig:6}(b).
By suitably adjusting the reflection or transmission coefficients of the \gls{ris} elements, the linear transformation experienced by the signals can be modified.
In practice, such reconfigurability is enabled by tunable microwave components embedded into the \gls{ris}, such as PIN diodes and varactors.
In addition to \glspl{ris}, other technologies exist to effectively reconfigure how \gls{rf} signals propagate from a transmitter to a receiver, such as \glspl{bd-ris}, \glspl{espar}, \glspl{dsa}, and reconfigurable antennas (including fluid and movable antennas).

Greater flexibility can be obtained by including multiple \glspl{ris} into the propagation environment, for instance by stacking several transmissive \glspl{ris}, as shown in Fig.~\ref{fig:6}(c).
This architecture, known as a \gls{sim} \cite{an23}, has been proposed as a means of realizing beamforming in the analog domain when placed close to the transmitting array.
A \gls{sim} exhibits an interesting structural resemblance to a \gls{nn}, as they both rely on cascaded layers that progressively transform the input signals.
However, this analogy breaks down when a \gls{sim} is composed only of linear microwave elements.
In that case, regardless of how sophisticated the resulting linear transform can be, the output signals remain a linear combination of the inputs.
To realize the functionality of a true \gls{nn} fully in the analog domain, nonlinearities would need to be incorporated within the layers, such as through diodes and amplifiers.

The computation of a \gls{milac} does not need to happen over a wireless channel, but it can also be carried out on a \gls{pcb}, as represented in Fig.~\ref{fig:6}(d).
A simple example is a fixed microwave circuit implemented in microstrip technology.
In microstrip circuits, metallic transmission lines are printed on a dielectric substrate to guide the microwave signals from the input ports to the output ports.
Therefore, these structures can be viewed as hardware implementations of linear transforms that are performed in the analog domain.
A well-known example is the Butler matrix, which maps the input signals onto predefined beam directions and thus realizes a \gls{dft}-like operation.
Beyond this operation, microstrip circuits can be designed to compute other linear operations, including the \gls{dft}, the Hadamard transform, and the Haar transform \cite{ner26}.
In Fig.~\ref{fig:dft}, we show a \gls{milac} prototype that computes the $4\times 4$ \gls{dft} operating at 2.4~GHz \cite{ner26}.

Reconfigurability can also be introduced in \gls{pcb} microwave networks by incorporating tunable components into the board.
A representative example is the analog transformation stage used in hybrid digital-analog beamforming \cite{soh16}, as shown in Fig.~\ref{fig:6}(e).
This network is typically implemented using power dividers, tunable phase shifters, and power combiners, which together realize a controllable linear transformation between the input and output signals.
By adjusting the phase shifters, the effective linear transformation implemented by the microwave network can be reconfigured, enabling adaptive beamforming with fewer \gls{rf} chains.
Beyond the use of phase shifters, alternative strategies that shape how \gls{rf} signals are radiated include \glspl{lma} and \glspl{dma}.

A natural question is what class of operations can be computed with the most general reconfigurable linear microwave network.
Maximum flexibility is obtained by interconnecting every port to all other ports, as well as to ground, through tunable impedance components \cite{ner25-1}, as illustrated in Fig.~\ref{fig:6}(f).
In this architecture, the tunable parameters of the network are the values of these impedance components, which can be implemented in practice using PIN diodes or varactors.
This architecture is especially important because its input-output relationship, while remaining linear in the input signal, is an interesting nonlinear function of the tunable parameters.
As a result, it can realize a much richer class of useful computations than simpler reconfigurable networks.
In particular, it is possible to identify a matrix $\mathbf{H}\in\mathbb{C}^{N\times M}$ and a scalar $\lambda$, both given by linear combinations of the tunable parameters, such that the computed transformation is $\mathbf{v}=\mathbf{H}^H(\mathbf{H}\mathbf{H}^H+\lambda\mathbf{I})^{-1}\mathbf{u}$ \cite{ner25-1}.
This operation, known as the \gls{lmmse} estimator, is of special interest in communications as it appears in \gls{r-zfbf} at the transmitter and in \gls{mmse} combining at the receiver \cite{ner25-2}.
Once the tunable impedance components have been configured according to a given $\mathbf{H}$ and $\lambda$, this operation is directly computed by the \gls{milac}, without requiring any digital computation of matrix inversions or matrix-matrix products.

In summary, linear microwave networks offer multiple ways to perform computation.
The analog computing can take place either over-the-air, through a wireless channel as in Fig.~\ref{fig:6}(a)-(c), or on-the-board, through a microwave circuit as in Fig.~\ref{fig:6}(d)-(f).
Moreover, these networks can be either fixed or reconfigurable.
Fixed networks, such as those in Fig.~\ref{fig:6}(a) and Fig.~\ref{fig:6}(d), implement a predetermined linear transformation.
By contrast, reconfigurable networks incorporate tunable components allowing to adapt the implemented linear transformation and, in some architectures, to realize input-output relationships whose dependence on the tunable parameters is interestingly nonlinear.

\subsection{Computational Complexity Advantages of MiLAC}
\label{sec:complexity}

\begin{table}[t]
\centering
\caption{Complexity of matrix operations performed with \\digital and analog computing.}
\begin{tabular}{@{}lll@{}}
\toprule
Operation              & Digital computing   & Analog computing\\
\midrule
Matrix-vector product  & $\mathcal{O}(MN)$   & $\mathcal{O}(1)$\\
LMMSE estimator        & $\mathcal{O}(MN^2)$ & $\mathcal{O}(MN)$\\
Matrix inversion       & $\mathcal{O}(N^3)$  & $\mathcal{O}(N^2)$\\
\bottomrule
\end{tabular}
\newline\newline
\label{tab}
\end{table}

An important feature of \gls{milac} is its ability to carry out matrix operations in the analog domain with significantly reduced computational complexity.
Table~\ref{tab} summarizes the complexity of representative operations computed digitally and with a \gls{milac}.

The most immediate example is the matrix-vector product.
Consider a fixed $M\times N$ matrix, physically implemented by a non-reconfigurable \gls{milac}.
Multiplying this matrix by an $N\times 1$ input vector requires a number of operations that scales as $\mathcal{O}(MN)$ in digital computing.
By contrast, when this transformation is implemented by a \gls{milac}, the multiplication is performed directly by wave propagation through the microwave network.
From a computational-complexity viewpoint, this corresponds to $\mathcal{O}(1)$, since the operation is carried out in a single analog propagation step and does not scale with $M$ or $N$.

Interestingly, more advanced operations can be computed efficiently when the \gls{milac} is implemented using interconnected tunable impedance components, as in Fig.~\ref{fig:6}(f).
It has been shown in \cite{ner25-1} that such a \gls{milac} can compute both the \gls{lmmse} estimator and matrix inversion by exploiting its input-output relationship given in Section~\ref{sec:6}.
In digital computing, evaluating the \gls{lmmse} estimator requires $\mathcal{O}(MN^2)$ operations, assuming $N\leq M$, while inverting an $N\times N$ matrix requires $\mathcal{O}(N^3)$ operations.
By contrast, when these two operations are performed by a \gls{milac}, their complexity is given by the cost of calculating the tunable parameters of the network for a given input data $\mathbf{H}$ and $\lambda$, or for a given matrix to be inverted.
Since the tunable parameters, namely the impedance values, are linear combinations of the input data, they can be computed with only $\mathcal{O}(MN)$ and $\mathcal{O}(N^2)$ operations for the \gls{lmmse} estimator and matrix inversion, respectively, as shown in \cite{ner25-1}.
Remarkably, although no digital algorithm is known to invert a matrix with quadratic complexity, this scaling can be achieved by a \gls{milac}.

\section{MiLAC-Aided Communications}

The computational capabilities of \gls{milac} make it attractive for beamforming in gigantic \gls{mimo} systems, where matrix operations could have prohibitive computational cost \cite{ner25-2}.
In this section, we introduce \gls{milac}-aided transmitting and receiving architectures, in which precoding and combining are carried out directly in the analog domain.
We also present advanced \gls{milac}-aided architectures that overcome the limitations of basic implementations and enable the same performance as fully digital beamforming in any communication system.

\subsection{MiLAC-Aided MIMO Architectures}

\begin{figure*}[t]
\centering
\includegraphics[width=0.71\textwidth]{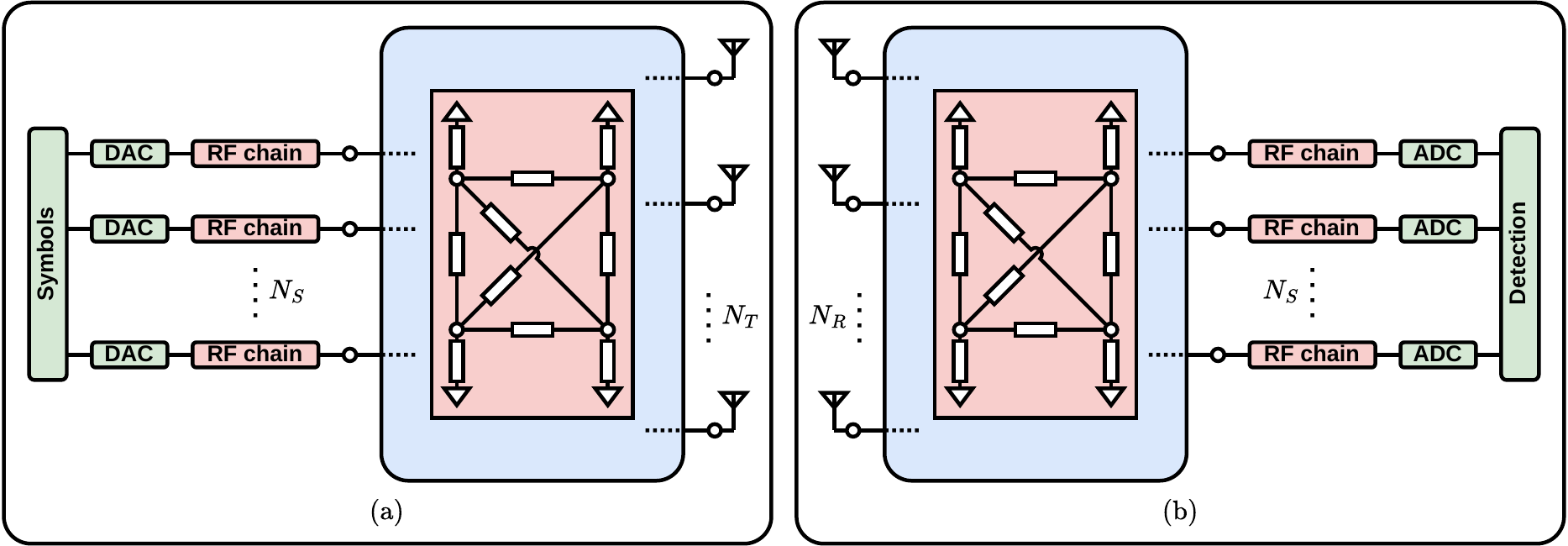}
\caption{(a) A MiLAC-aided transmitter precoding the symbols in the analog domain, and (b) a MiLAC-aided receiver combining the signals at the receiving antennas in the analog domain.
With lossless and reciprocal MiLACs, these architectures achieve the same performance as fully digital \gls{mimo} architectures in single-user \gls{mimo} systems.}
\label{fig:tx-rx}
\end{figure*}

At the transmitter, a \gls{milac} can be used to perform precoding directly in the analog domain, as illustrated in Fig.~\ref{fig:tx-rx}(a).
The information symbols are first generated according to a digital modulation scheme, such as $M$-\gls{qam}, and converted to analog signals by the \glspl{dac}.
These signals are then up-converted from baseband to the microwave domain by the \gls{rf} chains, which also carry out filtering and amplification.
The resulting microwave signals are fed to the \gls{milac}, which applies the desired linear transformation and produces the precoded signals to be radiated by the transmit antennas.
In this way, the beamforming operation is shifted from digital processing to the microwave network.

At the receiver, a \gls{milac} can be used to perform signal combining directly in the analog domain, as illustrated in Fig.~\ref{fig:tx-rx}(b).
The microwave signals collected by the receive antennas are first processed by the \gls{milac}, which combines them according to the desired linear transformation.
The combined outputs are then passed through the \gls{rf} chains, where they are down-converted from the microwave domain to baseband, and amplified and filtered appropriately.
Finally, the resulting analog baseband signals are digitized by the \glspl{adc} and used for symbol detection according to the adopted modulation scheme.

In practice, \gls{milac}-aided transmitters and receivers are often designed under the standard microwave constraints of losslessness and reciprocity.
Losslessness means that the microwave network neither dissipates nor amplifies power, while reciprocity means that its transmission characteristics are identical in both propagation directions.
These assumptions are aligned with practical implementations of passive microwave networks, which are typically made of reciprocal and (ideally) lossless components.
Under these constraints, it has been shown analytically that a \gls{milac}-aided transmitter or receiver can achieve the same performance as fully digital beamforming in single-user \gls{mimo} systems \cite{ner25-3}.
In multi-user systems, however, this optimality no longer holds in general, and is guaranteed only in the special case where the users experience mutually orthogonal channels \cite{fan26}.
Intuitively, this is because a lossless and reciprocal \gls{milac} that optimally transfers the power from the input to the output ports can only implement semi-unitary transformations.
This limitation motivates the development of enhanced \gls{milac}-aided \gls{mimo} architectures for more general multi-user systems.

\subsection{Advanced MiLAC-Aided MIMO Architectures}

\begin{figure*}[t]
\centering
\includegraphics[width=0.99\textwidth]{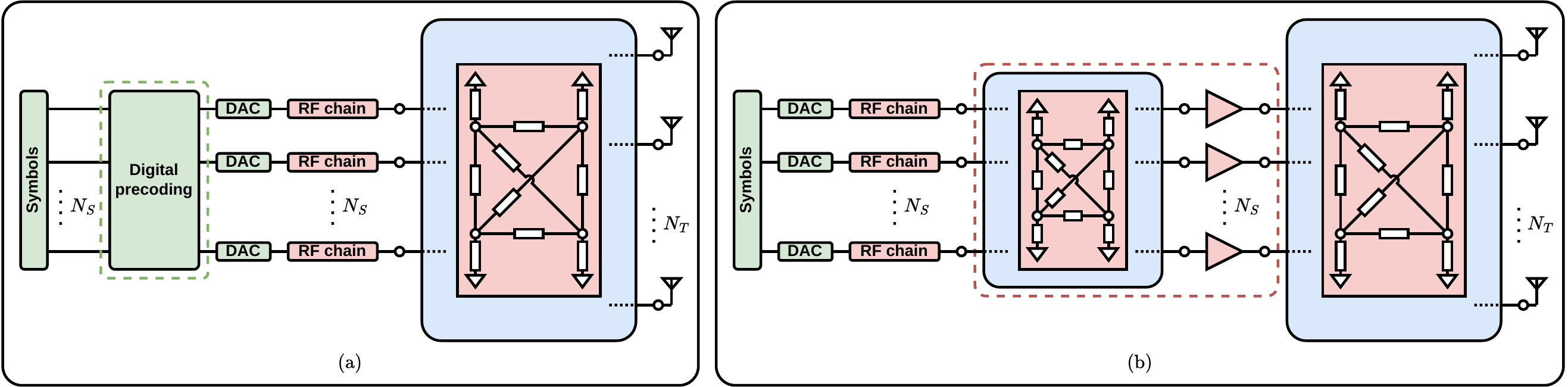}
\caption{(a) A hybrid digital-MiLAC-aided transmitter, and (b) a two-layer MiLAC-aided transmitter.
With lossless and reciprocal MiLACs, these architectures achieve the same performance as fully digital \gls{mimo} architectures in multi-user \gls{mimo} systems.}
\label{fig:advanced}
\end{figure*}

We now show that the \gls{milac}-aided \gls{mimo} architectures in Fig.~\ref{fig:tx-rx} can be enhanced to achieve the same flexibility as fully digital beamforming, even under the lossless and reciprocal constraints.
This can be done either by improving the digital processing or the analog processing capabilities of the proposed \gls{milac}-aided \gls{mimo} architectures.
In the following, we illustrate these two advanced architectures by focusing on the transmitter side, while the same discussion applies to the receiver as well.

A first approach is to complement the analog precoding performed by the \gls{milac} with additional digital precoding.
This leads to the hybrid digital-\gls{milac} architecture shown in Fig.~\ref{fig:advanced}(a), where the digital and analog domains can jointly realize an arbitrary precoding operation \cite{wu26}.
This can be intuitively explained by considering the \gls{svd} of the desired $N_T\times N_S$ precoding matrix.
The lossless and reciprocal \gls{milac} implements the semi-unitary transformation given by the first $N_S$ left singular vectors, while the remaining factors in the \gls{svd} are realized digitally.
As a result, hybrid digital-\gls{milac} beamforming can achieve optimal performance in general wireless communication scenarios using only as many \gls{rf} chains as data streams \cite{wu26}.

A second approach is to enhance the analog processing capabilities of \gls{milac}, without relying on digital precoding.
This can be achieved by cascading two \glspl{milac} with power amplifiers inserted between them, as illustrated in Fig.~\ref{fig:advanced}(b).
The resulting architecture, referred to as a two-layer \gls{milac}, can realize arbitrary precoding even when both \glspl{milac} are lossless and reciprocal \cite{zho26}.
This can again be understood from the \gls{svd} of the desired $N_T\times N_S$ precoding matrix.
The first \gls{milac} performs the semi-unitary transformation given by the $N_S$ right singular vectors, the gains of the power amplifiers are set as the $N_S$ singular values, and the second \gls{milac} performs the semi-unitary transformation given by the first $N_S$ left singular vectors.
In this way, this two-layer \gls{milac} architecture also achieves optimal performance with only as many \gls{rf} chains as transmitted streams, and without any digital precoding \cite{zho26}.

\section{Benefits of MiLAC for Communications}

We have shown how \glspl{milac} can realize high-performance beamforming directly in the analog domain by processing the same microwave signals used for communications, both at the transmitter and at the receiver.
In this section, we discuss the main benefits of these \gls{milac}-aided \gls{mimo} architectures.

\subsection{Low Number of RF Chains}

The first benefit of \gls{milac} for wireless communications is to drastically reduce the number of \gls{rf} chains.
In conventional fully digital \gls{mimo} architectures, each antenna requires a dedicated \gls{rf} chain, which includes expensive and power-hungry components such as local oscillators, mixers, and amplifiers.
Since the cost and power consumption scale with the array size, this approach is not scalable to massive and gigantic \gls{mimo}.
Hybrid digital-analog architectures were proposed to address this problem, and they only require a number of \gls{rf} chains that is twice the number of data streams \cite{soh16}.

To further reduce the number of \gls{rf} chains, \gls{milac}-aided architectures perform precoding and combining through the microwave network in the analog domain.
Therefore, the number of \gls{rf} chains only needs to be the same as the number of transmitted or received data streams.
Such a lower number of \gls{rf} chains naturally translates into reduced hardware cost and power consumption.
This advantage applies to the \gls{milac}-aided architectures in Fig.~\ref{fig:tx-rx}, as well as to the advanced architectures in Fig.~\ref{fig:advanced}.

\subsection{Low Resolution of ADCs and DACs}

A second benefit of \gls{milac} is that it only requires low-resolution \glspl{dac} at the transmitter and low-resolution \glspl{adc} at the receiver.
In conventional fully digital and hybrid digital-analog \gls{mimo} architectures, precoding and combining are carried out (at least in part) in the digital domain.
As a result, the signals at the \glspl{dac} and \glspl{adc} carry linear combinations of symbols, which typically require high-resolution quantization to avoid quantization noise and the resulting performance degradation.
However, the power consumption of \glspl{dac} and \glspl{adc} grows exponentially with their resolution, making low-resolution converters highly desirable.

In \gls{milac}-aided architectures, by contrast, beamforming is purely performed by the microwave network in the analog domain.
Therefore, the signals at the \glspl{dac} correspond directly to the transmitted symbols, while the signals at the \glspl{adc} are the combined signals used for symbol detection.
At the transmitter, since the symbols are drawn from constellations with finite cardinality, only a limited number of quantization levels is needed to represent them.
For example, with 4-\gls{qam}, the in-phase and quadrature components of the signals only take values in $\{-1,+1\}$, so 1-bit \glspl{dac} are sufficient.
Similarly, 16-\gls{qam} and 64-\gls{qam} require only 2-bit and 3-bit resolution, respectively.
Also at the receiver, \gls{milac} only requires low-resolution quantization because the \glspl{adc} operate directly on the signals that are quantized for symbol detection.
This advantage applies to the \gls{milac}-aided architectures in Fig.~\ref{fig:tx-rx} and to two-layer \gls{milac} architectures, but not to hybrid digital-\gls{milac} architectures, which rely on digital processing.

\subsection{Low Computational Complexity}

A third benefit of \gls{milac} is the reduction in computational complexity.
In conventional fully digital and hybrid digital-analog \gls{mimo} architectures, precoding at the transmitter and combining at the receiver must be performed digitally for every transmitted symbol vector.
Therefore, each symbol time requires the computation of a matrix-vector product with complexity $\mathcal{O}(N_TN_S)$.
Instead, in \gls{milac}-aided architectures, the same linear transformation is carried out directly by wave propagation through the microwave network.
Once the \gls{milac} has been configured, no digital matrix-vector product is required on a symbol-by-symbol basis.
This benefit does not hold for hybrid digital-\gls{milac} architectures, which still require $\mathcal{O}(N_S^2)$ computations at every symbol time.

The complexity reduction offered by \gls{milac} is even more significant in the case of \gls{r-zfbf} at the transmitter or \gls{mmse} combining at the receiver.
In digital implementations, these schemes require the evaluation of an \gls{lmmse}-like transformation at every channel coherence time, with complexity $\mathcal{O}(N_TN_S^2)$ because of the pseudo-inverse operation.
In a \gls{milac} implemented with interconnected tunable impedance components, however, this operation can be realized directly by the microwave network once the tunable components have been set appropriately.
The remaining digital computations are then limited to calculating these tunable components as a function of the channel matrix, whose complexity scales as $\mathcal{O}(N_TN_S)$ following Section~\ref{sec:complexity}.
Therefore, \gls{milac} not only removes the symbol-by-symbol matrix-vector computations, but can also reduce the complexity of popular beamforming strategies from cubic to quadratic.

\section{Challenges and Future Research}

\Gls{milac} offers significant opportunities for wireless communications, and naturally opens several important challenges and research directions.
In this section, we highlight some of the main open problems and promising avenues for future research.

\subsection{Hardware Impairments}

We have introduced \gls{milac} under the standard microwave assumptions of lossless and reciprocal tunable components.
While these assumptions enable a clean analytical characterization, practical \gls{rf} hardware is inevitably affected by non-idealities that must be accounted for to assess the true performance of \gls{milac}-aided systems.
First, tunable impedance components such as varactors and PIN diodes are often reconfigured within a discrete set of possible values, rather than with continuous values.
Second, practical components exhibit insertion losses, which may degrade the power delivered to the \gls{milac} output ports and therefore affect the beamforming performance.
Third, in \gls{milac}-aided \gls{mimo} systems, it is particularly important to deal with mutual coupling effects and impedance matching since beamforming is performed directly with the microwave signals \cite{xio26}.
A key research direction is therefore to develop accurate models for these hardware impairments and to design \gls{milac} optimization algorithms that account for them.

\subsection{Channel Estimation}

How to efficiently estimate the channel in \gls{milac}-aided systems is an important open problem.
Channel estimation is challenging in these systems because signal processing is carried out in the analog domain by the microwave network, which determines the observed signals available for digital processing.
A first study in this direction is \cite{zha26-1}, which shows that the computational capabilities of \gls{milac} can be exploited to obtain the \gls{mmse} channel estimate with low computational complexity.
These early results suggest that \gls{milac} can be used not only for beamforming, but also to accelerate key signal processing tasks at the physical layer.

\subsection{Wideband Communications}

Wideband communications constitute another important research direction for \gls{milac}-aided systems.
In narrowband systems, the microwave network implements a single linear transformation.
However, in wideband systems, the response of the \gls{milac} in general varies with the frequency.
Therefore, the wideband behavior of \gls{milac} must be accurately modeled to capture (and counteract or exploit) undesired effects such as beam squinting.
This model is necessary both to assess the performance limits of \gls{milac}-aided wideband systems and to develop optimization strategies that are reliable over the whole bandwidth of the signal.

\subsection{Low-Complexity MiLAC Architectures}

We have focused on \gls{milac} architectures where each port is interconnected to all others via a tunable impedance, which offer the greatest flexibility in the linear transformation that can be implemented.
The drawback of these architectures is their high circuit complexity, since the number of tunable components grows quadratically with the number of ports of the microwave network.
In practice, simpler \gls{milac} architectures can be designed, which include tunable impedance components only between selected pairs of ports, thereby reducing hardware complexity \cite{ner25-4}.
Understanding the resulting trade-off between circuit complexity and communication performance is an important direction for future research.
Interestingly, recent results show that very low-complexity \gls{milac} architectures can still be capacity-achieving in single-user \gls{mimo} systems \cite{ner25-4}.
This suggests that substantial hardware simplifications are possible without sacrificing optimality in some scenarios, and without considerable performance degradation in general \cite{zha26-2}.

\section{Conclusion}

Computing directly in the analog domain with communication signals is a promising way to relieve the growing digital burden in future wireless networks.
In this paper, we have introduced the concept of \gls{milac} and reviewed how linear microwave networks can shift key matrix computations from the digital baseband domain to the analog \gls{rf} domain.
By doing so, \glspl{milac} can enable scalable \gls{mimo} architectures with fewer \gls{rf} chains, lower-resolution \glspl{adc} and \glspl{dac}, and substantially reduced computational complexity.
As an emerging concept, \gls{milac} opens numerous avenues for future research in wireless communications.
Beyond communications, many fundamental questions on \gls{milac} also remain open, including the ultimate limits of its computational capabilities and its applications across the broader field of signal processing.

\bibliographystyle{IEEEtran}
\bibliography{IEEEabrv,main}

\section*{Biographies}

\noindent\textbf{Matteo Nerini} is a Postdoctoral Research Associate in the Department of Electrical and Electronic Engineering, Imperial College London, UK.\\

\noindent\textbf{Bruno Clerckx} is a Professor and the Head of the Communications and Signal Processing Group in the Department of Electrical and Electronic Engineering, Imperial College London, UK.

\end{document}